\documentclass[aps,prd,superscriptaddress,nofootinbib,floats,showpacs]{revtex4-2}
\usepackage{anyfontsize} 

\usepackage[normalem]{ulem}
\usepackage{xcolor}
\usepackage{float}
\usepackage{fancyhdr}
\usepackage{placeins}
\usepackage{amssymb}
\usepackage{stmaryrd}
\usepackage{amsfonts}
\usepackage{mathrsfs}
\usepackage{amsmath,amssymb,amsfonts}
\usepackage{graphicx}
\usepackage{subfigure}
\usepackage{color} 
\usepackage{fancyhdr}
\usepackage{hyperref} 
\def\be{\begin{equation}}
	\def\ee{\end{equation}}
\def\ba{\begin{eqnarray}}
	\def\ea{\end{eqnarray}}

\begin{document}

	\title{ Quantum Oppenheimer-Snyder models in loop quantum cosmology with Lorentz term}
	
	\author{Minyan Ou}
	\affiliation{Department of Physics, South China University of Technology, Guangzhou 510641, China}
	\author{ Xiangdong Zhang\footnote{Corresponding author. scxdzhang@scut.edu.cn}}
	\affiliation{Department of Physics, South China University of Technology, Guangzhou 510641, China}

	\begin{abstract}
		A novel quantum black hole model is derived by incorporating the Lorentzian term within the loop quantum cosmology framework of the quantum Oppenheimer-Snyder (qOS) model. This model features a quantum-corrected metric tensor, representing a deformation of the classical Schwarzschild solution. Investigations into the quasi-normal modes reveal that these quantum-corrected black holes exhibit stability against scalar perturbations. Notably, the exponential decay rate within the Lorentzian qOS model demonstrates a significant reduction compared to both the earlier qOS model devoid of this term and the standard Schwarzschild black hole. The higher overtones of the Lorentzian qOS model also differ significantly from those of the earlier qOS model and the standard Schwarzschild black hole, indicating that the near-horizon geometry is substantially modified. The thermodynamic properties with both positive and negative cosmological constant are considered. For the anti-de Sitter case, our analysis reveals that for small black hole masses, the temperature within this loop quantum gravity framework decreases as the mass diminishes, contrasting with classical black hole behavior. Furthermore, a logarithmic term emerges as the leading-order correction to the Bekenstein-Hawking entropy. Additionally, LQG corrections induce an extra phase transition in the black hole's heat capacity at smaller radius. While for the de Sitter case, the temperature increases as the mass decreases for small black holes, again differing from classical expectations. Similarly to the anti-de Sitter case, LQG corrections in the de Sitter case also lead to an extra phase transition in the heat capacity at small radius.
	\end{abstract}
	\maketitle

	\section{Introduction}\label{Intro}
	General Relativity (GR) encounters fundamental limitations when addressing spacetime singularities, particularly where curvature approaches Planck-scale magnitudes. This breakdown necessitates a quantum gravity framework capable of unifying quantum mechanics with GR to resolve these singularities. Among promising candidates, Loop Quantum Gravity (LQG) stands out due to its background-independent, nonperturbative nature and mathematical rigor \cite{thiemann2001introduction,ashtekar2004background,han2007fundamental,rovelli2004quantum}.
	
	Recent advancements in LQG have yielded significant breakthroughs, particularly in resolving the Big Bang singularity and elucidating Hawking-Bekenstein black hole entropy \cite{song2022thermodynamics,rovelli1996black,meissner2004black}. The application of LQG principles to homogeneous, isotropic universes has established Loop Quantum Cosmology (LQC), whose hallmark achievement is replacing the classical Big Bang singularity with a quantum bounce mechanism \cite{bojowald2008loop,ashtekar2011loop,agullo2014loop,ashtekar2006quantum,ashtekar2006quantumanalytical,ashtekar2006quantumdynamics,ashtekar2008robustness,yang2009alternative,zhang2016higher}.

	In classical gravitational theory, the Oppenheimer-Snyder (OS) model \cite{oppenheimer1939continued} provides a foundational framework for studying gravitational collapse through pressureless, homogeneous dust dynamics governed by Friedmann equations. While valuable, this model inherits GR's singularity problem. Recent work has addressed this limitation by applying LQC corrections to the OS framework, resulting in the quantum Oppenheimer-Snyder (qOS) model \cite{lewandowski2023quantum}, which has subsequently undergone extensive theoretical investigation \cite{shi2024higher,gong2024quasinormal,yang2023shadow,zhang2023loop}.
	
	Within the current qOS framework, the interior dynamics of black holes are predominantly analyzed using the simplest Loop Quantum Cosmology (LQC) approach, specifically the Ashtekar-Pawlowski-Singh (APS) model \cite{ashtekar2006quantum}. It is noteworthy that the Hamiltonian constraint in Loop Quantum Gravity (LQG) comprises both Euclidean and Lorentzian components. Classically, these terms exhibit proportionality, yet the APS model employs a pre-quantization reduction that retains only the Euclidean term \cite{ashtekar2011loop}. This simplification yields a symmetric bounce behavior during both pre- and post-bounce phases \cite{ashtekar2006quantum}.
	
	Emerging research indicates that this conventional treatment fails to fully capture LQG's complete feature set. Crucially, full theory quantization demonstrates distinct behaviors between Euclidean and Lorentzian terms \cite{assanioussi2018emergent,assanioussi2019emergent,han2020effective,yang2023alternative}. Through Thiemann's regularization technique, a refined LQC model incorporating the Lorentzian term has been developed. This advancement introduces significant modifications to the Friedmann equation \cite{long2021energy,li2018towards,zhang2021loop} manifesting an emergent de-Sitter epoch and asymmetric bounce dynamics. This raises a pivotal question: Would substituting the APS model with this Lorentzian-enhanced LQC preserve the qOS framework's core conclusions? Our investigation addresses this through systematic derivation and analysis the physical properties of the modified qOS-LQC model.

	One important physical property is that the black hole perturbation dynamics: When subjected to external perturbations, black holes exhibit a distinctive gravitational wave emission pattern that unfolds in three sequential stages. The initial burst phase, whose intensity and duration depend on the perturbation conditions, is followed by the quasinormal mode (QNM) ringdown - a hallmark feature where spacetime oscillations decay with characteristic complex frequencies \cite{lin2024quasinormal}. These QNM signatures, recently confirmed through groundbreaking gravitational wave detections \cite{abbott2017gw170817,abbott2017gravitational} have become powerful tools for probing the fundamental geometry of black holes. The ringdown profile's precision is governed by both the fundamental mode ($n=0$) and its higher-order overtones ($n > 0$) \cite{giesler2019black,oshita2021ease,forteza2021high} with each overtone contributing unique information about the black hole's structure. The final phase, known as late-time tail evolution, displays either power-law or exponential decay patterns that encode crucial information about the black hole's properties \cite{shao2024scalar}, particularly regarding the nature of its Cauchy horizon \cite{cardoso2018quasinormal,poisson1990internal}.
	Consequently, the QNMs and the late-time tail, are applied to test the stability of black holes and encode important information about the black holes. To this end, we employ the finite element method and WKB approximation to investigate the QNMs, overtone, and late-time tails of scalar perturbations in quantum black holes.
	
	Recently, black hole thermodynamics has attracted great interest. The Hawking radiation paradigm \cite{hawking1974black,hawking1975particle}, established black holes as thermodynamic entities possessing temperature and entropy. The four laws of black hole thermodynamics \cite{bardeen1973four}, parallel classical thermodynamic principles, with Bekenstein-Hawking entropy bridging classical and quantum gravitational theories \cite{bekenstein1973black}.
	
	Concurrently, cosmological observations of universal acceleration necessitate dark energy models, where the cosmological constant $\Lambda$ remains the predominant explanation \cite{weinberg1989cosmological,peebles2003cosmological}. In particular, the recent cosmological observations of the Universe are consistent with a de Sitter (dS) spacetime, characterized by a positive cosmological constant ($\Lambda > 0$). In this context, the physical and thermodynamic properties of de Sitter black holes have also been extensively studied \cite{du2023topological,sekiwa2006thermodynamics}.  Meanwhile, the success of the AdS/CFT correspondence has highlighted the theoretical importance of anti-de Sitter (AdS) spacetime, characterized by a negative cosmological constant ($\Lambda < 0$). The AdS black holes exhibit unique thermodynamic behaviors, such as phase transitions and critical phenomena \cite{kubizvnak2012p,dayyani2018critical,wei2020extended}, which are not observed in asymptotically flat spacetimes. One of the most interesting black hole phase transitions is the Hawking-Page phase transition \cite{hawking1983thermodynamics}. In this work, we focus on investigating the thermodynamic properties in both de Sitter and anti-de Sitter spacetimes. 
	
	This paper is organized as follows. Firstly, we derive the new model by combining the modified Friedmann equation with the qOS model, and analyze its physical properties in \text{Sect. II}. Then, we investigate the QNMs, the
	late-time tail behavior, and the behavior of higher overtones in \text{Sect. III}. In the \text{Sect. IV}, we examine the thermodynamic properties of the model.
	Finally, we summarize and discuss our main results in \text{Sect. V}. Throughout this work, we use geometrical units in which $G = c = \hbar = 1$.

	\section{The qOS model in LQC with Lorentz term}\label{SandW1}
	The qOS model's spacetime comprises the pressureless dust ball and the vacuum region outside the dustball, the metric tensor inside of the dustball can be described by the FRW metric as
	\begin{eqnarray}
		ds_{in}^2 = -d\tau^2 + a(\tau)^2( d\tilde{r}^2 + \tilde{r}^2 d\Omega^2),
		\label{eq: inside}
	\end{eqnarray}
	where \(d\Omega^2 = d\theta^2 + \sin^2\theta \ d\phi^2\), and \(a(\tau)\) is the scale factor of the universe. The original qOS model uses the Ashtekar-Pawlowski-Singh (APS) model \cite{ashtekar2006quantum} deciping. However, the APS version of LQC model only consider the Euclidean term in Hamiltonian constraint and stand for a simplified model. The recent advances of LQC found that the Lorentz term should also need to be considered and will bring the significant changes for the cosmic evolution such as the de-Sitter epoch will emerged. The dynamics of the \(a(\tau)\) of LQC with Lorentz term governed by the modified Friedmann equation \cite{long2021energy,li2018towards}
	
	\begin{equation}
		H^2 = (\frac{\dot{a}}{a})^2 = \frac{1}{\gamma^2 \Delta} f(\rho) (1-f(\rho)) \left(1-\frac{\rho}{\rho_{c}}\right),
		\label{eq: modified Fr equation}
	\end{equation}
	where \[f(\rho) = \frac{1 \pm \sqrt{1-\frac{\rho}{\rho_{c}}}}{2 (1 + \gamma^2)}, \qquad  \rho = \frac{M}{\frac{4}{3} \pi \tilde{r}^3 a^3}.\]
	
	Here \(\Delta = 2 \pi \sqrt{3} \gamma G \hbar \), which is the smallest non-vanishing area eigenvalue from the full theory \cite{thiemann2001introduction,ashtekar2004background}.
	The $\gamma$ is the Barbero-Immirzi parameter. Using the black hole thermodynamics in LQG, so we could set the value being
	0.2375. \({\rho_{c}} =  \frac{3}{ 32 \pi G \Delta \gamma^2 (1 + \gamma^2)}\)  is the critical energy density of this
	model. 
	
	We should note that the modified Friedmann equation possesses two distinct branches. The first branch is \(f(\rho)_{+} = \frac{1 + \sqrt{1-\frac{\rho}{\rho_{c}}}}{2 (1 + \gamma^2)}\),
	though it has no cosmological constant added at the classical evolution. However, its quantum geometric effects are equivalent to a positive effective cosmological constant of Planckian order \cite{assanioussi2018emergent}.
	The effective cosmological constant is $\Lambda_{eff} = \frac{3}{(1+\gamma^2)^2 \Delta}$, resulting in an de-Sitter epoch. When \(\rho \ll \rho_{c}\), the $f(\rho)_{+}$ branch of the modified Friedmann equation reduces to
	
	\begin{eqnarray}
		H^2_{+} = \frac{8 \pi G_{\alpha}}{3} \rho + \frac{\Lambda_{eff}}{3}.
		\label{eq: H2 +}
	\end{eqnarray}
	where \(G_{\alpha} = \frac{1 - 5 \gamma^{2}}{\gamma^{2} + 1} G\). This differs from the classical case, as it includes both a positive effective cosmological constant and a modified Newton’s constant.
	
	The second branch is given by \(f(\rho)_{-} = \frac{1 - \sqrt{1-\frac{\rho}{\rho_{c}}}}{2 (1 + \gamma^2)}\). When \(\rho \ll \rho_{c}\), it reduces to
	
	\begin{eqnarray}
		H^2_{-} = \frac{8 \pi G}{3} \rho. 
		\label{eq: H2 -}
	\end{eqnarray}
	This branch corresponds to a classical universe in which no effective cosmological constant appears.
	
	Outside the dust ball's region, we assume that the spacetime is spherically symmetric and static, so the metric reads
	
	\begin{eqnarray}
		ds_{out}^2 = -F(r)dt^2 + G(r)^{-1} dr^2 + d\Omega^2.
		\label{eq: outside metric}
	\end{eqnarray}
	
	   It is worthy noting that in the cosmological case, the \( f(\rho)_{+} \) branch is not suitable for describing our universe, and therefore the bounce proceeds from the \( f(\rho)_{+} \) branch to the \( f(\rho)_{-} \) branch.
		Conversely, in the collapse case, the \( f(\rho)_{-} \) branch provides a more appropriate description of stellar collapse, so the bounce proceeds from the \( f(\rho)_{-} \) branch to the \( f(\rho)_{+} \) branch also sounds.
		Since in both scenarios we are primarily interested in post-bounce branch, the resulting exterior solutions differ accordingly. A detailed explanation will be provided below.

	\subsection{The vacuum exterior for the $f(\rho)_{-}$ branch}\label{MPD}
	
	As Eq. (\ref{eq: H2 +}) shows, the emergence of the effective cosmological constant and the modified Newton's constant is not consistent with current cosmological observations. Therefore, it is not suitable for describing our expanding universe.
	In the cosmological scenario, the bounce transitions from the $f(\rho)_{+}$ branch to the $f(\rho)_{-}$ branch, where the universe corresponding to the $f(\rho)_{-}$ branch is the one we inhabit. Accordingly, we focus on the $f(\rho)_{-}$ in the following analysis.
	
	Having fixed the branch of the modified Friedmann equation, we can now determine the metric outside the dust ball region. To determine the unknown functions $F(r)$ and $G(r)$, we need to consider the Darmois-Israel junction condition \cite{darmois1927equations,israel1966singular}.
	We first consider the radial coordinate of the surface of the dust ball, denoted by $\tilde{r_{0}}$. Then, the hypersurface induced from Eq.(\ref{eq: inside}) is
	
	\begin{eqnarray}
		ds_{in}^2|_{\Sigma} = - d\tau^2 + a(\tau)^2 \tilde{r_{0}}^2 d\Omega^2.
		\label{eq: inside hypersurface}
	\end{eqnarray}
	
	We assume that every point on the surface of the dust ball shares the same proper time $\tau$, which evolves as the collapse proceeds. For convenience, we use $\tau$ to parameterize the hypersurface in the vacuum region outside the dust ball.
	Then, we have
	
	\begin{eqnarray}
		ds_{out}^2|_{\Sigma} = -\bigg(F\big(r(\tau)\big) \dot{t}(\tau)^2 - G\big(r(\tau)\big)^{-1} \dot{r}(\tau)^2 \bigg) d\tau^2 + r(\tau)^2 d\Omega^2.
		\label{eq: outside hypersurface}
	\end{eqnarray}
	
	To construct the complete spacetime, we match the interior region of the dust ball to the exterior region of a spherically symmetric, static spacetime using the Darmois–Israel junction conditions. The first junction condition requires the metric
	to be continuous across the boundary. Consequently, at the boundary, the metric components satisfy
	
	\begin{eqnarray}
		1 = \bigg(F\big(r(\tau)\big) \dot{t}(\tau)^2 - G\big(r(\tau)\big)^{-1} \dot{r}(\tau)^2 \bigg), \quad  \quad a(\tau) = r(\tau).
		\label{eq: metic equal}
	\end{eqnarray}
	
	According to the second Darmois–Israel junction condition, the extrinsic curvature must be continuous across the boundary. In the spacetime of the interior dust ball, the non-vanishing components are
	
	\begin{eqnarray}
		K^{in}_{\theta \theta} = a(\tau) \tilde{r_{0}}, \quad K^{in}_{\phi \phi} = a(\tau) \tilde{r_{0}} \sin^2\theta .
		\label{eq: inside extrinsic curvature}
	\end{eqnarray}
	For the exterior vacuum spacetime, the non-zero components of the extrinsic curvature are
	
	\begin{eqnarray}
		K^{out}_{\theta \theta} = r F(r(\tau)) \dot{t} \sqrt{F(r(\tau))^{-1} G(r(\tau))}, \quad K^{out}_{\phi \phi} = r F(r(\tau)) \dot{t} \sqrt{F(r(\tau))^{-1} G(r(\tau))} \sin^2\theta .
		\label{eq: outside extrinsic curvature}
	\end{eqnarray}
	
	Since the exterior spacetime is assumed to be static, $\xi^{a} = (\frac{\partial}{\partial t})^{a}$ is a Killing vector, which allows us to define the conserved energy
	
	\begin{eqnarray}
			\xi_{\mu} u^{\mu} = g^{out}_{\mu \nu} \xi^{\nu} u^{\mu} = F\big(r(\tau)\big)  \dot{t}(\tau) = -E, 
		\label{eq: conserved energy}
	\end{eqnarray}
	where $ u^{\mu} = (\dot{t},\dot{r},0,0)$.
	
	Combining Eq. (\ref{eq: inside extrinsic curvature}) and Eq. (\ref{eq: outside extrinsic curvature}), and using Eq. (\ref{eq: conserved energy}), we obtain
	
	\begin{eqnarray}
		a(\tau) \tilde{r_{0}} =  r F(r(\tau)) \dot{t}(\tau) \sqrt{F(r(\tau))^{-1} G(r(\tau))} = r E \sqrt{F(r(\tau))^{-1} G(r(\tau))}.
		\label{eq: extrinsic curvature equal}
	\end{eqnarray}
	Using Eq. (\ref{eq: extrinsic curvature equal}) together with the metric continuity condition Eq. (\ref{eq: metic equal}), we find
	
	\begin{eqnarray}
		F(r(\tau)) = E^2 G(r(\tau)), \quad G(r(\tau)) = 1 - \dot{r}^2.
		\label{eq: junction condition}
	\end{eqnarray}
	Combining Eq. (\ref{eq: junction condition}) with the modified Friedmann equation Eq. (\ref{eq: modified Fr equation}), we obtain
	
	\begin{eqnarray}
		G(r(\tau)) = 1 - H^2 r^2 = 1 - \frac{r^2}{\gamma^2 \Delta} f(\rho) \big(1-f(\rho)\big) \big(1-\frac{\rho}{\rho_{c}}\big) .
		\label{eq: G(r)}
	\end{eqnarray}
	For simplicity and without loss generality, we can set $E = 1$, which gives $F(r) =  G(r)$. The exterior metric then takes the form
	
		\begin{align}
			ds_{out}^2 &= - F(r) dt^2 + F(r)^{-1} dr^2 + d\Omega^2,\nonumber\\
			F(r) &= 1 - \frac{
				r^{2}}{
				2\gamma^{2}(1+\gamma^{2})\Delta
			}\left(1 - \dfrac{1 - \sqrt{1 - \dfrac{3M}{4\pi r^{3}\rho_{c}}}}{2(1+\gamma^{2})}\right)
			\left(1 - \sqrt{1 - \dfrac{3M}{4\pi r^{3}\rho_{c}}}\right)
			\left(1 - \dfrac{3M}{4\pi r^{3}\rho_{c}}\right).
			\label{eq: new metric -}
		\end{align}

	It should be noted that the metric (\ref{eq: new metric -}) is asymptotically flat, and is valid for 
	
	\begin{eqnarray}
		r \geq r_{b} = (\frac{3 M}{4 \pi \rho_{c}})^{\frac{1}{3}}.
		\label{eq rb}
	\end{eqnarray}
	$r_{b}$ is the dust ball minimal radial.
	
\begin{figure}[H]
		\centering
		\includegraphics[width=0.5\textwidth]{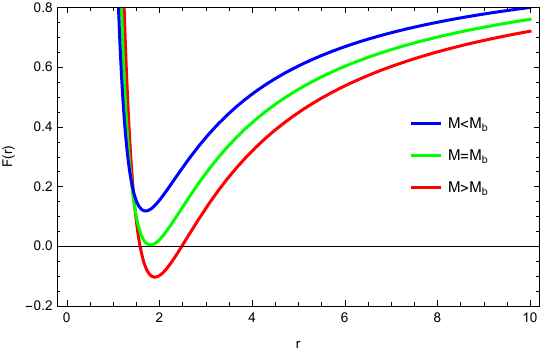}
		\caption{\small The metric function $F(r)$ versus r with different M.}
		\label{frimage}
	\end{figure}

	In Figure \ref{frimage}, we investigate the effect of the mass of black hole $M$ on $F(r)$ while keeping the other parameters fixed. For simplicity, in our calculations,
	we set the remaining parameters as $\gamma =0.2375 $. As shown in the figure, when $M$ is bigger than the extreme value $M_{b}$, the metric function $F(r)$ has two roots,
	corresponding to the two horizons of the black hole. When $M = M_{b}$, $F(r)$ has exactly one root, indicating the presence of a single horizon. When $M < M_{b}$,
	the black hole has no horizon. The critical value is found to be $M_{b} = 1.2015$. For generality, in this work, we only consider the case of black holes with two horizons.
	
	\FloatBarrier
	
	\subsection{The vacuum exterior for the $f(\rho)_{+}$ branch }\label{MPD}
	
	In the classical Oppenheimer–Snyder (OS) model, a homogeneous dust ball collapses inevitably to a singularity. In contrast, in the de Sitter OS model \cite{markovic2000gravitational} contracts to a minimal radius and undergoes a bounce, followed by expansion. 
	Studying the physical properties and formation thresholds of primordial black holes formed from gravitational collapse within the LQC framework provides a valuable probe of the quantum nature of gravity and its potential connection to dark matter \cite{papanikolaou2023primordial}.
	Therefore, considering gravitational collapse in this context is both important and meaningful.
	
	In the context of stellar collapse, the interior of the dust ball is described by the metric
		\begin{align}
			ds_{-}^2 = -d\tau^2 + a(\tau)^2( d\tilde{r}^2 + \tilde{r}^2 d\Omega^2),
		\end{align}
		The dynamics of \(a(\tau)\) are governed by the $f(\rho)_{-}$ branch of the modified Friedmann equation, given by
		\begin{align}
			H_{-}^2 = (\frac{\dot{a}}{a})^2 = \frac{1}{
				2\gamma^{2}(1+\gamma^{2})\Delta
			}
			\left(1 - \dfrac{1 - \sqrt{1 - \dfrac{3M}{4\pi r^{3}\rho_{c}}}}{2(1+\gamma^{2})}\right)
			\left(1 - \sqrt{1 - \dfrac{3M}{4\pi r^{3}\rho_{c}}}\right)
			\left(1 - \dfrac{3M}{4\pi r^{3}\rho_{c}}\right).
			\label{eq: inside of collapse}
	\end{align}
	The dust ball starts from a nearly static configuration at large $r$, contracts gradually, and crosses the past event horizon. It continues collapsing until it reaches the minimal radius $r_{b}$ where the bounce occurs.
	After the bounce, the evolution transitions to the $f(\rho)_{+}$ branch, representing the expanding phase, described by the metric
		\begin{align}
			ds_{+}^2 = -d\tau^2 + a(\tau)^2( d\tilde{r}^2 + \tilde{r}^2 d\Omega^2),
		\end{align}
		with the scale factor \(a(\tau)\) governed by the $f(\rho)_{+}$ branch:
		\begin{align}
			H_{+}^2 = (\frac{\dot{a}}{a})^2 = \frac{1}{
				2\gamma^{2}(1+\gamma^{2})\Delta}
			\left(1 - \dfrac{1 + \sqrt{1 - \dfrac{3M}{4\pi r^{3}\rho_{c}}}}{2(1+\gamma^{2})}\right)
			\left(1 + \sqrt{1 - \dfrac{3M}{4\pi r^{3}\rho_{c}}}\right)
			\left(1 - \dfrac{3M}{4\pi r^{3}\rho_{c}}\right).
			\label{eq: outside of collapse}
		\end{align}
	
	The dust ball then expands outward, crosses the future cosmological horizon, and continues to infinity. 
	
	It should be noted that only the $f(\rho)_{-}$ branch reproduces the classical Oppenheimer–Snyder collapse in the \(\rho \ll \rho_{c}\), while the $f(\rho)_{+}$ branch does not, as discussed previously. Consequently, the big bounce corresponds to a transition from the $f(\rho)_{-}$ branch to the $f(\rho)_{+}$ branch, and we are primarily interested in post-bounce branch. 
	
	Once the appropriate branch of the modified Friedmann equation is identified, the exterior metric of the dust ball can be determined following the same procedure as outlined before, resulting in the metric
			
		\begin{align}
			ds_{out}^2 &= - F(r) dt^2 + F(r)^{-1} dr^2 + d\Omega^2,\nonumber\\
			F(r) &= 1 - \frac{
				r^{2}}{
				2 \gamma^{2} (1+\gamma^{2}) \Delta
			}
			\left( 
			1 - \frac{ 1 + \sqrt{ 1 - \frac{3M}{4\pi r^{3}\rho_{c}} } }
			{ 2(1+\gamma^{2}) }
			\right)
			\left( 1 + \sqrt{ 1 - \frac{3M}{4\pi r^{3}\rho_{c}} } \right)
			\left( 1 - \frac{3M}{4\pi r^{3}\rho_{c}} \right).
			\label{eq: new metric +}
		\end{align}

	\section{QUASINORMAL MODES}\label{SandW2}
	
	 As mentioned earlier, the vacuum exterior of the $f(\rho)_{-}$ branch corresponds to a spacetime containing a black hole, whereas the $f(\rho)_{+}$ branch admits only a cosmological horizon. Moreover, the vacuum exterior of the $f(\rho)_{-}$ branch appears to be more consistent with cosmological observations. Therefore, in this section, we restrict our analysis to the $f(\rho)_{-}$ branch.

	\subsection{Massless scalar and the effective potential}\label{MPD1}
	
	To gain a deeper understanding of the spacetime properties of the metric (\ref{eq: new metric -}), now we consider a massless scalar field perturbation in the above background. The massless scalar field $ \boldsymbol\Psi $ is governed by the Klein-Gordon (KG) equation:
	
	\begin{eqnarray}
		\Box \boldsymbol\Psi = \frac{1}{\sqrt{- g}} \partial_{\mu} (\sqrt{- g} g^{\mu \nu} \partial_{\nu} \boldsymbol\Psi(t, r, \theta, \phi)) = 0.
		\label{eq: KG}
	\end{eqnarray}
	
	For the background metric (\ref{eq: new metric -}) is static and spherically symmetric, we can separate the massless scalar field $\boldsymbol\Psi$ in the form
	
	\begin{eqnarray}
		\boldsymbol\Psi(t, r, \theta, \phi) = \sum\limits_{l, m} \frac{\Phi (t, r)}{r} Y_{l, m} (\theta, \phi), 
		\label{eq: separate psi}
	\end{eqnarray}
	where $ Y_{l, m} $ is the spherical harmonics, and $l$ and $m$ denote the multipole number and azimuthal number respectively. Substituting Eq. (\ref{eq: separate psi}) into
	Eq. (\ref{eq: KG}), the scalar equation Eq. (\ref{eq: KG}) reduces to
	
	\begin{align}
		- \frac{\partial^{2} \Phi(t, r)}{\partial t^{2}} + F(r)^2 \frac{\partial^{2} \Phi(t, r)}{\partial r^{2}} + F(r) F(r)^\prime \frac{\partial \Phi(t, r)}{\partial r} 
		- \big(\frac{l (l + 1)}{r^2} F(r) + \frac{F(r) F(r)^\prime}{r} \big) \Phi(t, r) = 0.
		\label{eq: complex KG}
	\end{align}
	In order to map the radial region to $(- \infty, + \infty)$, we define the tortoise radial coordinate $r_{*}$ by
	\begin{eqnarray}
		d r_{*} = \frac{d r}{F(r)}.
	\end{eqnarray}
	Then Eq. (\ref{eq: complex KG}) can be casted as  
	\begin{eqnarray}
		- \frac{\partial^2 \Phi}{\partial t^2} + \frac{\partial^2 \Phi}{\partial r_{*}^2} - V(r) \Phi = 0,
		\label{eq: effective KG}
	\end{eqnarray}
	Assuming that $\Phi(t, r) = \Psi(r) e^{- i \omega t}$, we yield
	\begin{eqnarray}
		\frac{\partial^2 \Psi}{\partial r_{*}^2} + (\omega^2 - V(r)) \Psi = 0.
	\end{eqnarray}
	where the $\omega$ is the frequency of QNMs, and $V(r)$ is the effective potential which given by
	\begin{eqnarray}
		V(r) = F(r)(\frac{l (l + 1)}{r^2} + \frac{F(r)^\prime}{r}).
	\end{eqnarray}
	Physically, in the event horizon, nothing can escape, so it only have incoming waves. And the metric Eq.(\ref{eq: new metric}) is asymptotically flat, there are no incoming 
	waves from infinity. Thus the boundary conditions can be chosen as
	\begin{eqnarray}
		\Psi(r) \approx e^{- i \omega r_{*}} (r \rightarrow r_{h}), \qquad  \Psi(r) \approx e^{ i \omega r_{*}} (r \rightarrow + \infty).
		\label{boundary conditions}
	\end{eqnarray}

	\subsection{The finite element method}\label{MPD}
	
	In this work, we will introduce the finite element method and use it to solve the Eq. (\ref{eq: effective KG}) to obtain the time-domain profile of $\Phi(t, r)$. 
	We discretize the coordinates $ t = i \triangle t $ and $ r_{*} = j \triangle r_{*} $, where the $i$ and $j$ are integers. The scalar field $\Phi$ and
	the effective potential $ V $ are also discretized:
	\begin{eqnarray}
		\Phi(t, r) = \Phi(i \triangle t, j \triangle r_{*}) \equiv \Phi(i, j), \qquad   V(r_{*}) = V(j \triangle r_{*}) \equiv V(j).
	\end{eqnarray}
	So, the Eq.(\ref{eq: effective KG}) becomes a set of iterative algebraic equations:
	\begin{align}
		- \frac{\Phi(i + 1, j) - 2 \Phi(i, j) + \Phi(i - 1, j)}{\triangle t^2} + \frac{\Phi(i, j + 1) - 2 \Phi(i, j) + \Phi(i, j - 1)}{\triangle r_{*}^2}
		- V(j) \Phi(i, j) = 0.
		\label{eq: separate effective KG}
	\end{align}
	We can rewritten the Eq.(\ref{eq: separate effective KG}) as 
	\begin{align}
		\Phi(i + 1, j) = - \Phi(i -1, j) + (2 - 2 \frac{\triangle t^2}{\triangle r_{*}^2} - \triangle t^2 V(j)) \Phi(i, j) + \frac{\triangle t^2}{\triangle r_{*}^2} 
		(\Phi(i, j + 1) + \Phi(i, j - 1)).
	\end{align}
	The von Neumann stability condition require $ \frac{\triangle t}{\triangle r_{*}} < 1$, so we set $ \frac{\triangle t}{\triangle r_{*}} =\frac{1}{2} $ in our 
	numerical calculations. We consider the initial Gaussian distribution:
	\begin{eqnarray}
		\Phi(t = 0, r_{*}) = e^{- \frac{(r_{*} - \overline{a})^2}{2}}, \qquad and \qquad \Phi(t < 0, r_{*}) = 0. 
	\end{eqnarray}
	the value of $\overline{a}$ will be chosen accordingly.
	
	\subsection{WKB approximation}\label{MPD2}
	
	The WKB approximation is a semi-analytic method widely used to compute the QNMs of black holes. B. Schutz and C. Will first applied the first-order WKB method
	to study black hole scattering problems \cite{schutz1985black}. Subsequently, S. Iyer and C. Will extended the method to third order \cite{iyer1987black}, resulting
	in improved accuracy, while Konoplya further pushed it to sixth order, achieving even greater precision \cite{konoplya2003quasinormal}. More recently, by employing
	Padé approximants, Matyjasek and Opala substantially enhanced the accuracy of the WKB method and extended it up to the 13th order \cite{matyjasek2017quasinormal}. 
	Then Konoplya and collaborators further improved the WKB method, significantly enhancing its precision \cite{konoplya2019higher}.
	
	The WKB approximation is used to solve wave-like equations where the effective potential has a barrier-like shape (a single peak) and approaches a constant as
	\( r^* \to \pm\infty \). The main idea of this method is to match the approximate solutions expanded near the asymptotic regions—both near the horizon and at infinity
	via the two turning points, using Taylor expansions around the top of the potential barrier. Finally, for spherically symmetric backgrounds with an effective potential
	$V(r)$ that does not depend on the frequency $\omega$, the WKB formula can be expressed as \cite{konoplya2019higher}:
	
	\begin{align}
		\omega^{2} = & V_{0} + A_{2}(\mathcal{K}^{2}) + A_{4}(\mathcal{K}^{2}) + A_{6}(\mathcal{K}^{2}) + \ldots \nonumber\\
		& - \mathrm{i} \mathcal{K} \sqrt{-2 V_{2}}(1 + A_{3}(\mathcal{K}^{2}) + A_{5}(\mathcal{K}^{2}) + \ldots),
	\end{align}
	
	where \[\mathcal{K} = n + \frac{1}{2}, \quad n = 0,1,2,3,\ldots \]
	
	In this expression, $n$ labels the overtone number, $V_{0}$ denotes the maximum of the effective potential $V(x)$, and $V_{n}$ refers to the $n$th derivative of $V(x)$ evaluated
	at this maximum. $A_{n}(\mathcal{K}^{2})$ is the $n$th-order correction, expressed as a polynomial in $\mathcal{K}^{2}$ with rational coefficients.

	\subsection{The QNMs results and the late-time tail}\label{MPD3}
	
	In this subsection, we investigate the angular momentum $l$ and the mass $M$ how to affect the effective potential and QNMs under scalar perturbations, and further analyze the behavior of the late-time tail.
	In the final stage of the scalar field evolution, the late-time tail becomes dominant. And the inverse power-law tail usually arises in the asymptotically flat spacetime. 
	
\begin{figure}[htbp]
		\centering
		\begin{minipage}[t]{0.45\textwidth}
			\centering
			\includegraphics[width=\textwidth]{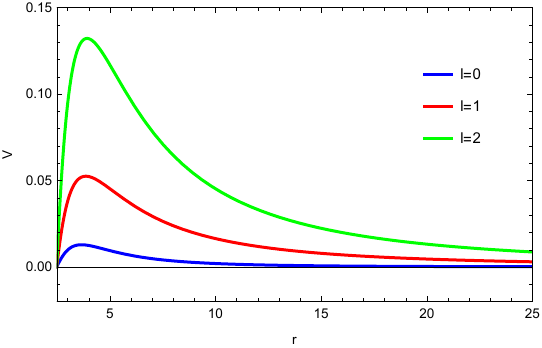}
			\caption{\small The effective potential $V(r)$ for different angular momentum $l$. Parameters used: $M=4.4$.}
			\label{VLimage}
		\end{minipage}
		\hfill
		\begin{minipage}[t]{0.45\textwidth}
			\centering
			\includegraphics[width=\textwidth]{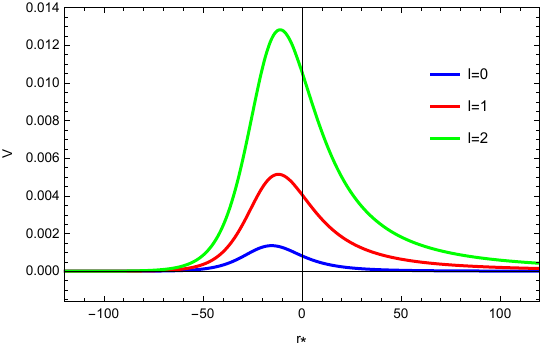}
			\caption{\small The effective potential $V(r_{*})$ for different angular momentum $l$. Parameters used: $M=4.4$.}
			\label{VXLimage}
		\end{minipage}
	\end{figure}

\begin{figure}[htbp]
		\centering
		\begin{minipage}[t]{0.45\textwidth}
			\centering
			\includegraphics[width=\textwidth]{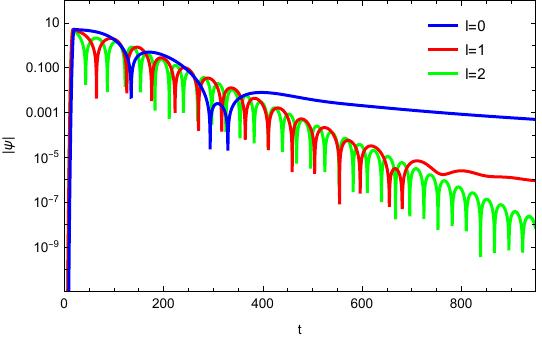}
			\caption{\small The time evolution of the scalar field $\Phi$ for different angular momentum $l$. Parameters used: $M=4.4$.}
			\label{VLpsiimage}
		\end{minipage}
		\hfill
		\begin{minipage}[t]{0.45\textwidth}
			\centering
			\includegraphics[width=\textwidth]{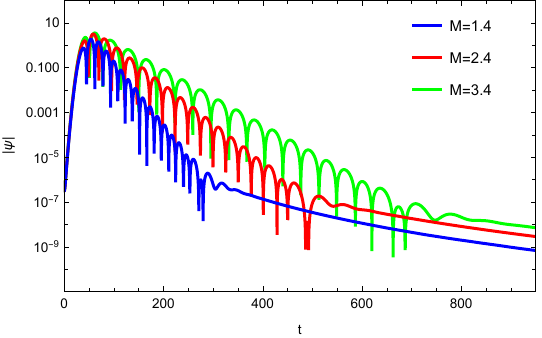}
			\caption{\small The time evolution of the scalar field $\Phi$ for different mass $M$ with $l=1$.}
			\label{VMpsiimage}
		\end{minipage}
	\end{figure}

	Using the finite element method to solve equation (\ref{eq: separate effective KG}), we obtain the time-domain profile.
	In Fig. \ref{VLimage} and Fig. \ref{VXLimage}, we can see that when we use the scalar waves with different angular momentum $l$ to perturb the same black hole,
	the height of the effective potential changes. When angular momentum $l$ increases, the peak of the effective potential increases. We should note that the 
	tortoise radial coordinate $r_{*}$ doesn't change the shape of the effective potential, it just stech it. The time-domain profiles correspond to this effective potential are
	shown in Fig. \ref{VLpsiimage}. As shown in the figure, we can note that the radiative decay follows the inverse power-law at late times. And when angular momentum $l$ increases,
	the quasinormal frequencies become longer-lived and dominate the time-domain signal.
	
\begin{figure}[htbp]
		\centering
		\begin{minipage}[t]{0.45\textwidth}
			\centering
			\includegraphics[width=\textwidth]{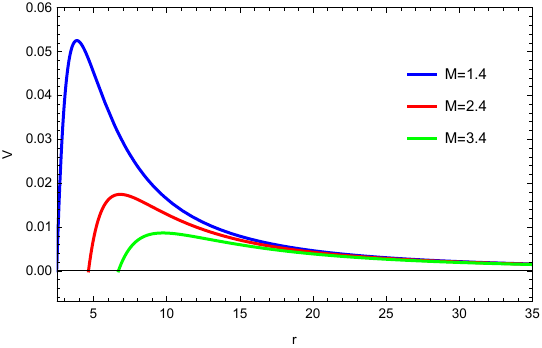}
			\caption{\small The effective potential $V(r)$ for different mass $M$. Parameters used: $l=1$.}
			\label{VMimage}
		\end{minipage}
		\hfill
		\begin{minipage}[t]{0.45\textwidth}
			\centering
			\includegraphics[width=\textwidth]{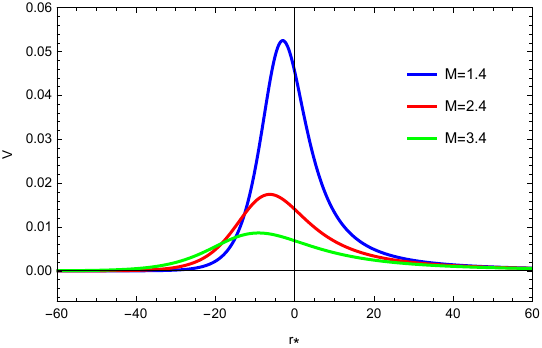}
			\caption{\small The effective potential $V(r_{*})$ for different mass $M$. Parameters used: $l=1$.}
			\label{VXMimage}
		\end{minipage}
	\end{figure}

	In Fig. \ref{VMimage} and Fig. \ref{VXMimage}, we can see that when the mass of black hole $M$ increases, the peak of the effective potential decreases. The time-domain profiles correspond
	to this effective potentials shown in Figure \ref{VMpsiimage}. In the figure, we observe that as the mass of the LQG-inspired black hole increases, the exponential decay rate decreases and the late-time
	tail appears later. Furthermore, the late-time tails corresponding to different black hole masses all follow an inverse power-law behavior, indicating that the black hole
	mass has little effect on the late-time tail.
	
\begin{figure}[htbp]
		\centering
		\begin{minipage}[t]{0.45\textwidth}
			\centering
			\includegraphics[width=\textwidth]{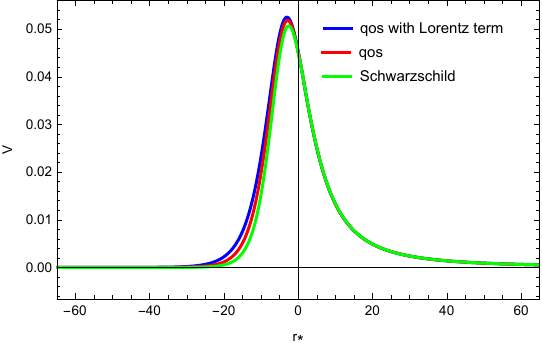}
			\caption{\small The effective potential $V(r_{*})$ for different black hole models: Schwarzschild, qOS, and qOS with Lorentz term. Parameters used: $M=1.4$, $l=1$, $\overset{\sim}{\alpha} = 1.6$, where $\overset{\sim}{\alpha}$ denotes the quantum correction parameter in the qOS model.}
			\label{Vrimage}
		\end{minipage}
		\hfill
		\begin{minipage}[t]{0.45\textwidth}
			\centering
			\includegraphics[width=\textwidth]{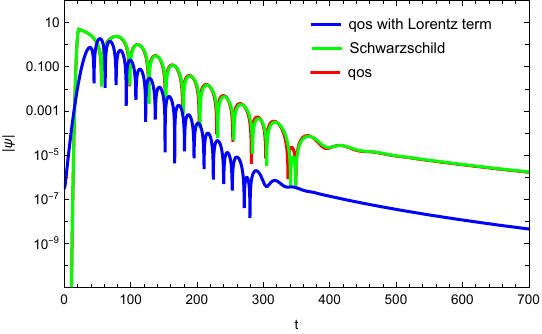}
			\caption{\small The time evolution of the scalar field $\Phi$ for different black hole models: Schwarzschild, qOS, and qOS with Lorentz term. Parameters used: $M=1.4$, $l=1$, $\overset{\sim}{\alpha} = 1.6$, where $\overset{\sim}{\alpha}$ denotes the quantum correction parameter in the qOS model.}
			\label{threemodelpsiimag}
		\end{minipage}
	\end{figure}

	Fig. \ref{Vrimage} presents the effective potential $V(r_{*})$ for the Schwarzschild black hole, the black hole in the qOS model, and the qOS model with Lorentz term respectively,
	all evaluated with the same mass parameter $M$ and angular momentum $l$. Near the event horizon, the effective potential exhibits slight deformations, indicating modifications in the spacetime structure
	of the event horizon. To further explore these structural differences, we analyze the time-domain profiles in Fig. \ref{threemodelpsiimag}. As indicated in the Figure, the profile of the black hole in the previous qOS model
	closely matches that of the Schwarzschild black hole, whereas the exponential decay rate in the qOS model with the Lorentz term shows a
	noticeable decrease. 
	It should be noted that the deformation induced by quantum effects is minimal compared to the classical case.
	The profiles of both the Schwarzschild black hole and the black hole in the qOS model with Lorentz term display a late-time inverse power-law tail,
	indicating that the quantum effects of the qOS model with Lorentz term have little impact on the late-time tail behavior.

	\subsection{Higher overtones}\label{MPD4}
	
	While the fundamental mode usually dominates the ringdown phase, it is insufficient for accurately extracting the mass and spin of the remnant black hole formed from a binary merger. This limitation
	can be addressed by including the first few overtones \cite{giesler2019black}. Analysis of LIGO’s observational data from the GW150914 gravitational-wave signal \cite{abbott2016binary} provides evidence for the presence of the first overtone.
	Although corrections to Einstein’s theory must modify the geometry near the event horizon, quasinormal modes (QNMs) are generally believed to be insensitive to the near-horizon structure, as the dominant QNMs are determined primarily
	by the region near the peak of the effective potential. However, to better probe the geometry near the event horizon and thereby gain deeper insight into quantum gravity, it is necessary to consider the first few overtones, which are particularly
	sensitive to the geometric structure around the event horizon \cite{konoplya2024first}. In certain scenarios, these overtones can be significantly excited and may even be detectable during the early ringdown phase by LISA \cite{oshita2023slowly}.
	
	To further investigate the differences in the spacetime structure among the Schwarzschild black hole, the qOS model, and the qOS model with the Lorentz term, we analyze their higher overtones using the WKB method.
	As shown in Table \ref{Table1} and Table \ref{Table2}, the three models share a common trend: the real part of the frequency decreases, while the imaginary part increases with increasing overtone number $n$.
	Although the overall trend is consistent, significant differences in the overtone frequencies are observed for the qOS model with the Lorentz term compared to the Schwarzschild black hole, which means that $Re(\omega)$ and $Im(\omega)$
	experience an outburst. Some quantum-corrected black holes, higher-derivative gravity models, and the RN black hole exhibit similar behavior \cite{konoplya2023quasinormal,berti2003asymptotic,konoplya2023bardeen}. We also find that the overtone
	frequencies of the qOS model with the Lorentz term show significant differences from those of the qOS model, which means that the Lorentz term has an effect on the structure of the event horizon, and this effect is reflected in the higher overtones.
	
\begin{figure}[htbp]
		\centering
		\begin{minipage}[t]{0.45\textwidth}
			\centering
			\includegraphics[width=\textwidth]{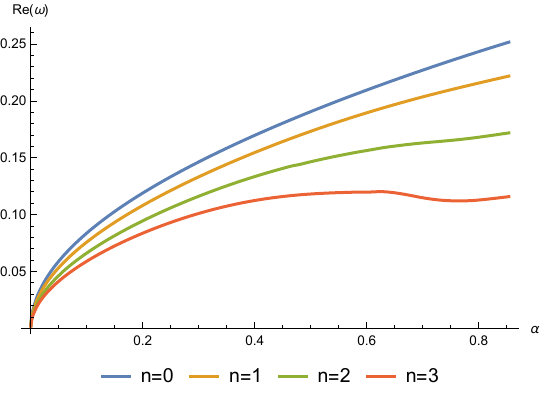}
			\caption{\small $Re(\omega)$ as a function of $\alpha$ for the black hole of qOS model with Lorentz term for $l=1$.}
			\label{l1reimage}
		\end{minipage}
		\hfill
		\begin{minipage}[t]{0.45\textwidth}
			\centering
			\includegraphics[width=\textwidth]{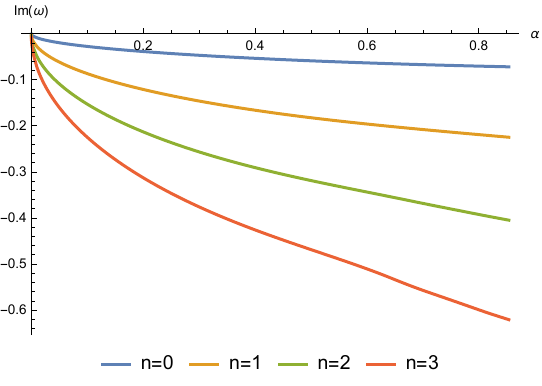}
			\caption{\small $Im(\omega)$ as a function of $\alpha$ for the black hole of qOS model with Lorentz term for $l=1$.}
			\label{l1imimage}
		\end{minipage}
	\end{figure}

\begin{figure}[htbp]
		\centering
		\begin{minipage}[t]{0.45\textwidth}
			\centering
			\includegraphics[width=\textwidth]{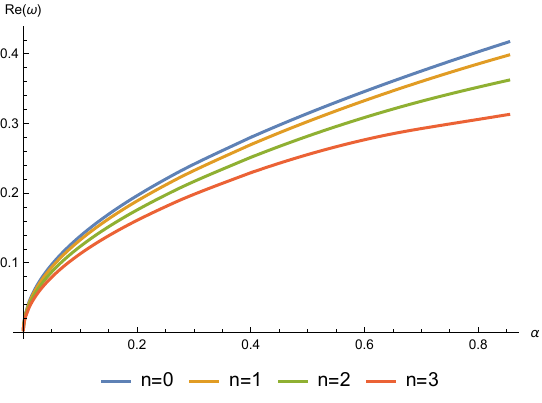}
			\caption{\small $Re(\omega)$ as a function of $\alpha$ for the black hole of qOS model with Lorentz term for $l=2$.}
			\label{l2reimage}
		\end{minipage}
		\hfill
		\begin{minipage}[t]{0.45\textwidth}
			\centering
			\includegraphics[width=\textwidth]{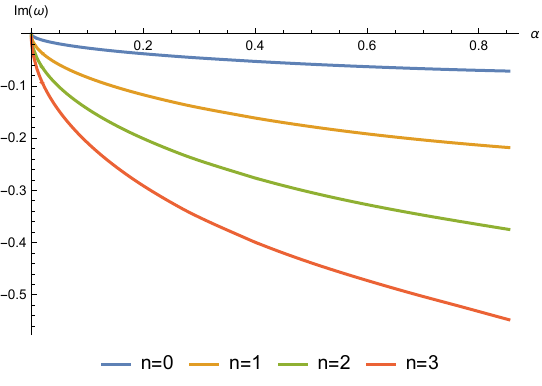}
			\caption{\small $Im(\omega)$ as a function of $\alpha$ for the black hole of qOS model with Lorentz term for $l=2$.}
			\label{l2imimage}
		\end{minipage}
	\end{figure}

	To this aim, in order to gain a better understanding how the qOS model with the Lorentz term affects the spacetime structure, we analyze how the quantum-corrected parameter $\alpha$ influences the higher overtones. Here, we set $\alpha = \frac{8 \gamma^{2} (1 + \gamma^{2}) \Delta}{M^{2}}$,
	which is a dimensionless parameter. The maximum value of the quantum-corrected parameter $\alpha$ is determined by the extremal value $M_{b}$. In Fig. \ref{l1reimage} and Fig. \ref{l2reimage}, we observe that for $l =1$ and $l =2$, $Re(\omega)$ 
	increases monotonically with increasing $\alpha$. However, in Fig. \ref{l1reimage}, the $Re(\omega)$ of $n=3$ exhibits a non-monotonic behavior. In Fig. \ref{l1imimage} and Fig. \ref{l2imimage}, we observe that for both $l =1$ and $l =2$, $Im(\omega)$
	decreases monotonically as $\alpha$ increases. 
	
	\FloatBarrier
	
	\begin{table}[htbp]
		\centering
		\caption{The dominant mode and first four overtones for three different black hole models: Schwarzschild, qOS, and qOS with Lorentz term. Parameters used: $M=1.4$,
			$l=1$, $\overset{\sim}{\alpha} = 1.6$, where $\overset{\sim}{\alpha}$ denotes the quantum correction parameter in the qOS model. }
		\label{Table1}
		\begin{tabular}{|c|c|c|c|}
			\hline
			$n$ & qOS with Lorentz term & Schwarzschild  &  qOS  \\
			\hline
			0 & 0.214719$-$0.064490i & 0.209239$-$0.069754i & 0.212885$-$0.066801i \\
			\hline
			1 & 0.193770 $-$0.199489i & 0.188793$-$0.218719i & 0.193923$-$0.207754i \\
			\hline
			2 & 0.158299$-$0.350276i & 0.163875$-$0.386053i & 0.168338$-$0.363045i \\
			\hline
			3 & 0.119553$-$0.523653i & 0.144981$-$0.562810i & 0.140515$-$0.525821i \\
			\hline
			4 & 0.095652$-$0.716769i & 0.130806$-$0.743082i & 0.109629$-$0.700130i \\
			\hline
		\end{tabular}
		\label{tab:qnm_final}
	\end{table}

	\begin{table}[H]
		\centering
		\caption{The dominant mode and first four overtones for three different black hole models: Schwarzschild, qOS, and qOS with Lorentz term. Parameters used: $M=1.4$,
			$l=2$, $\overset{\sim}{\alpha} = 1.6$, where $\overset{\sim}{\alpha}$ denotes the quantum correction parameter in the qOS model.}
		\label{Table2}
		\begin{tabular}{|c|c|c|c|}
			\hline
			$n$ & qOS with Lorentz term & Schwarzschild & qOS model \\
			\hline
			0 & 0.354598$-$0.064223i & 0.345460$-$0.069113i & 0.351235$-$0.0663599i \\
			\hline
			1 & 0.340954 $-$0.195116i & 0.331322$-$0.211146i & 0.338155$-$0.202051i \\
			\hline
			2 & 0.315535$-$0.332964i & 0.307538$-$0.363254i & 0.315317$-$0.345657i \\
			\hline
			3 & 0.281461$-$0.481173i & 0.281330$-$0.527130i & 0.287958$-$0.498779i \\
			\hline
			4 & 0.242022$-$0.641936i & 0.257775$-$0.699845i & 0.259685$-$0.659395i \\
			\hline
		\end{tabular}
		\label{tab:qnm_final}
	\end{table}

	\FloatBarrier
	
	\section{THERMODYNAMIC PROPERTIES}\label{SandW3}	
	
	In the preceding sections, we derived the metric corresponding to the post-bounce branch of the spacetime without a cosmological constant for both the cosmological and collapse scenarios, and analyzed its properties.
	As discussed earlier, the metric of the spacetime on the post-bounce branch differs between the cosmological and collapse cases. In the cosmological case, the vacuum exterior metric corresponding to the post-bounce branch is denoted by $f(\rho)_{-}$, whose geometry asymptotically
	approaches flat spacetime. In contrast, for the collapse case, the vacuum exterior metric corresponding to the post-bounce branch is given by $f(\rho)_{+}$, which asymptotically approaches a de Sitter spacetime with an effective cosmological constant \cite{assanioussi2018emergent,zhang2021loop}.
	
	However, our Universe is currently undergoing an accelerated expansion, the simplest possible explanation of which is the presence of a positive cosmological constant ($\Lambda > 0$). This motivates us to investigate the effect of the
	cosmological constant on the spacetime, as it introduces important modifications to its structure and dynamics. Moreover, the success of the AdS/CFT correspondence has drawn considerable attention to the case with a negative cosmological constant ($\Lambda < 0$).
	Introducing a negative cosmological constant in black hole physics also leads to richer phenomena.
	Hence, the inclusion of a cosmological constant in the solution have significance both in the practical and the theoretical sense.
	To investigate the thermodynamic properties of the vacuum exterior metrics for the $f(\rho)_-$ and $f(\rho)_+$ branches, it is necessary to extend the metrics to include the cosmological constant term \cite{lin2024effective,wang2024thermodynamics}.
	By substituting the previous Friedmann equation Eq. (\ref{eq: modified Fr equation}) with the modified Friedmann equation \cite{bentivegna2008anti} that incorporates the cosmological constant, and following the same derivation process as in the preceding
	sections, we can derive the metric that includes the cosmological constant, which is
	\begin{eqnarray}\label{eq: metric with cosmological constant}
		ds^2 = -F(r) dt^2 + F(r)^{-1} dr^2 + d\Omega^2,
		\label{metric of lamla}
	\end{eqnarray}
	
	where \[F(r) = 1 - \frac{r^2}{\gamma^2 \Delta} f(\rho) \Big(1-f(\rho)\Big)(1-\frac{\rho}{\rho_{c}}), \qquad  f(\rho)_{\pm} = \frac{1 \pm \sqrt{1-\frac{\rho}{\rho_{c}}}}{2 (1 + \gamma^2)}, \qquad  \rho = \frac{M}{\frac{4}{3} \pi r^3} + \frac{\Lambda}{8 \pi G}.\]
	
		To capture the richer thermodynamic features of the black hole, we incorporate a negative cosmological constant into the vacuum exterior metric of the $f(\rho)_-$ branch.
		As discussed earlier, the metric of the $f(\rho)_+$ branch does not contain a cosmological constant, but its quantum geometric effects are equivalent to a positive effective cosmological constant, and resulting in an de-Sitter epoch.
	To examine how the quantum geometric effects influence the thermodynamic properties, we compare the thermodynamic behavior of this spacetime with that of the classical case. Furthermore, as we show below, incorporating a negative cosmological constant into the metric of this
	case does not alter our main conclusions.

	\subsection{Anti-de Sitter spacetime}\label{MPD} 
	
	Due to the rich and intricate thermodynamic properties exhibited by AdS spacetimes, this subsection will primarily focus on the metric of the $f(\rho)_-$ branch with a negative cosmological constant.
	In particular, we will discuss the thermodynamic properties in this context, including the Hawking temperature, heat capacity, and entropy, which are central to understanding the behavior
	of black holes and phase transitions in AdS backgrounds.
	
\begin{figure}[!htb]
		\centering
		\begin{minipage}[t]{0.45\textwidth}
			\centering
			\includegraphics[width=\textwidth]{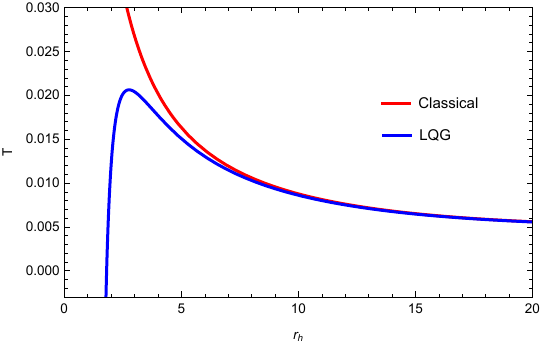}
			\caption{\small Hawking temperature of the LQG and Schwarzschild black holes as a function of even horizon $r_{h}$. Parameters used: $\Lambda = -0.001$.  }
			\label{AdsTimage}
		\end{minipage}
		\hfill
		\begin{minipage}[t]{0.45\textwidth}
			\centering
			\includegraphics[width=\textwidth]{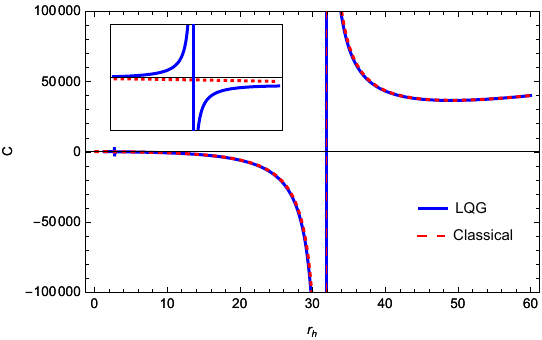}
			\caption{\small Heat capacity of the LQG and Schwarzschild black holes as a function of even horizon $r_{h}$. Parameters used: $\Lambda = -0.001$.  }
			\label{AdsCimage}
		\end{minipage}
	\end{figure}

	The Hawking temperature, which describes the thermal radiation emitted by the black hole, is related to the surface gravity at the event horizon. Specifically, the Hawking temperature
	can be expressed as:
	\begin{eqnarray}
		T_{H} = \frac{\mathcal{K}}{2 \pi} = \frac{1}{4 \pi} \frac{\partial F(r)}{\partial r} \Bigg|_{r = r_{h}}.
	\end{eqnarray}
	where $\mathcal{K}$ is the surface gravity of the black hole, $r_{h}$ is the event horizon radius, which is the largest root of the Eq. (\ref{eq: metric with cosmological constant}) with the
	$f(\rho)_{-}$ branch. In Fig. \ref{AdsTimage}, we compare the
	Hawking temperature for two types of black holes: the LQG-inspired black hole and the classical counterpart.
	As shown in the graph, the temperature monotonically decreases as $r_{h}$ increases in the classical scenario. When  $r_{h} $ approaches zero, the temperature diverges, tending towards infinity.
	In contrast, the LQG-inspired black holes exhibits a different behavior. 
	The temperature starts at the extremal case, where the $f(\rho)_{-}$ branch of the metric in Eq.~(\ref{metric of lamla}) has a single horizon. Initially, the temperature
		increases as $r_{h}$ grows, reaching a maximum value. Afterward, the temperature decreases, eventually approaching the same behavior as the classical black hole at large $r_{h}$.
	In some modified gravity theories, the Hawking temperature exhibits a similar behavior \cite{lin2024effective,huang2023novel}.
	
	In the AdS space, a large black hole is thermodynamically stable when it has a positive heat capacity, and it's instability when it has a negative heat capacity. So, the heat capacity of
	a black hole provides insight into the stability of black hole. The heat capacity of black hole \cite{cai2002gauss} is:
	\begin{eqnarray}
		C = \big(\frac{\partial M}{\partial T}) = \big(\frac{\partial M}{\partial r_{h}}) / \big( \frac{\partial T}{\partial r_{h}}).
	\end{eqnarray}
	From the Eq.(\ref{eq: metric with cosmological constant}) with the $f(\rho)_{-}$ branch, we can use the event horizon $r_{h}$ to express the black hole's mass. In Fig.\ref{AdsCimage},
	we plot the curve of the heat capacity for $r_{h}$ and compare
	it with the classical scenario. As shown in the graph, we can find that the LQG-inspired black hole compared with the classical scenario has an extra phase transition. Similar phenomena can be observed
	in some modified gravity \cite{lin2024effective}.
	
\begin{figure}[!htb]
		\centering
		\includegraphics[width=0.5\textwidth]{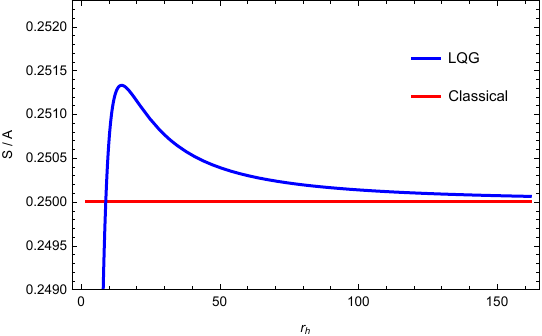}
		\caption{\small The ratio $\frac{S}{A}$ of the LQG and Schwarzschild black holes as a function of even horizon $r_{h}$. Parameters used: $\Lambda = -0.001$.}
		\label{AdsKimage}
	\end{figure}

	In AdS spacetime, the negative cosmological constant is interpreted as a positive thermodynamic pressure $P$ \cite{kubizvnak2015black,kubizvnak2017blackLambda}, with its conjugate quantity being the thermodynamic volume $V$. Specifically, the
	pressure is related to the cosmological constant by: \( P = \frac{- \Lambda}{8 \pi}\). Thus the first law of black hole thermodynamics takes the form \cite{dolan2011pressure}: 
	\ba
	dU = TdS - PdV,
	\ea where the $U$ is the internal energy, 
	$T$ is the Hawking temperature, $S$ denotes the entropy. From the first law, the entropy $S$ can be derived as follows:
	\begin{eqnarray}
		S = \int \frac{dM}{T}.
	\end{eqnarray}
	In Fig. \ref{AdsKimage}, we display the behavior of the ratio $\frac{S}{A}$ as a function of the horizon radius $r_{h}$ and compare the LQG-inspired black hole with the classical
	scenario, where $A$ is the area of the event horizon. As shown in the graph, we can see that the ratio $\frac{S}{A}$ for the LQG-inspired black hole exhibits a non-monotonic behavior: it initially increases with $r_{h}$,
	then decreases, and eventually approaches the classical result at large $r_{h}$.
	
	In general, the entropy of a classical black hole satisfy the area law, \(S = \frac{A}{4}\). However, in cases involving quantum corrections and higher-dimensional gravity, the area law no longer
	holds. Compared to the general case, an additional logarithmic correction term appears \cite{diaz2012isolated,frodden2012black,lin2024effective}. In our work, the entropy also contains a logarithmic correction term, which is:
	\begin{eqnarray}
		S = \frac{A}{4} + 1.506 \Delta \log A.
	\end{eqnarray}
	
	\subsection{de Sitter spacetime}\label{MPD} 
	
	The metric of the $f(\rho)_+$ branch does not contain a cosmological constant, but its quantum geometric effects are equivalent to a positive
		effective cosmological constant of Planckian order \cite{assanioussi2018emergent}. The effective cosmological constant is $\Lambda_{eff} = \frac{3}{(1+\gamma^2)^2 \Delta}$, resulting in an de-Sitter epoch.

	To investigate how quantum geometric effects influence its thermodynamics, we compare this solution with the de Sitter Schwarzschild black hole,
	which has the same effective cosmological constant $\Lambda_{eff}$. For the $f(\rho)_{+}$ branch of the Eq. (\ref{eq: metric with cosmological constant}), we define the cosmological horizon as $r_{c}$.
	
\begin{figure}[!htb]
		\centering
		\begin{minipage}[t]{0.45\textwidth}
			\centering
			\includegraphics[width=\textwidth]{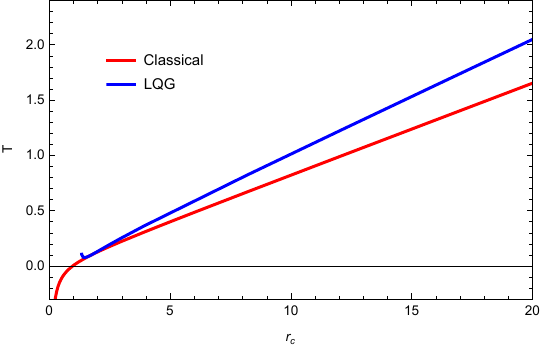}
			\caption{\small Hawking temperature of the LQG and the classical Schwarzschild-dS black holes as a function of cosmological horizon $r_{c}$.}
			\label{dsTimage}
		\end{minipage}
		\hfill
		\begin{minipage}[t]{0.45\textwidth}
			\centering
			\includegraphics[width=\textwidth]{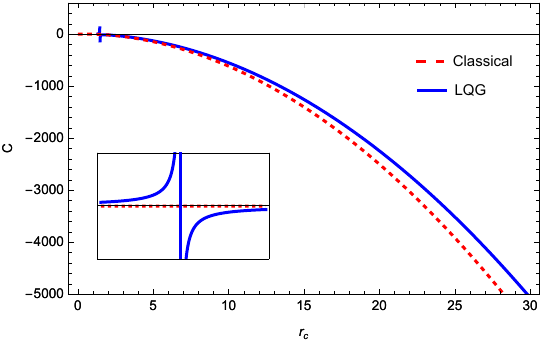}
			\caption{\small Heat capacity of the LQG and the classical Schwarzschild-dS black holes as a function of cosmological horizon $r_{c}$.} 
			\label{dsCimage}
		\end{minipage}
	\end{figure}

	In Fig. \ref{dsTimage}, we investigate the Hawking temperature how to effect of the cosmological horizon. In the de Sitter Schwarzschild black hole, the Hawking temperature increases linearly with $r_{c}$.
	In contrast, for the LQG-inspired black hole, the Hawking temperature first decreases with increasing $r_{c}$, and then increases linearly. Moreover, the temperature increases at different
	rates in the two cases, indicating that although the temperature profiles have a similar overall shape, their thermodynamic behaviors may not be identical. It should be noted that
	the Hawking temperature is usually positive, therefore, the temperature of the classical black hole starts from zero.
	
	We also investigate the relationship of the heat capture with the cosmological horizon $r_{c}$ on the Fig. \ref{dsCimage}. We can see that, compared to the de Sitter Schwarzschild black hole, the LQG solution has an additional phase transition, which reflects the influence of quantum effects on its thermodynamic properties. At large horizon radius $r_{c}$, the heat capacity of both the de Sitter Schwarzschild black hole and the LQG solution becomes negative, indicating that both of them are thermodynamically unstable.
	
	We can also incorporate a negative cosmological constant into this metric, while ensuring that its absolute value does not exceed the effective cosmological constant $\Lambda_{eff}$. We find that the conclusion remains unchanged.

	\section{ CONCLUSIONS AND DISCUSSIONS}\label{SandW4}
	
	This study systematically explores the qOS framework within Loop Quantum Cosmology (LQC) augmented by a Lorentz term, establishing its efficacy in characterizing quantum-gravitational dynamics. The model exhibits an asymmetric quantum bounce bifurcated into two phases: the $f(\rho)_+$ branch governing de Sitter epoch, and the $f(\rho)_-$ branch asymptotic flat. In the cosmological case, the bounce allow to transit from $f(\rho)_+$ branch to $f(\rho)_-$ branch, whereas in the collapse scenario,it transitions from $f(\rho)_-$ branch to $f(\rho)_+$ branch. In both cases, the post-bounce branch describes the universe in which we currently live. Consequently, our investigation focuses on the phenomenologically relevant post-bounce sector. 
	
	In the cosmological case, the vacuum exterior associated with the post-bounce \(f(\rho)_{-}\) branch contains a black hole, whereas in the collapse scenario, the vacuum exterior corresponding to the post-bounce \(f(\rho)_{+}\) branch features only a cosmological horizon. Therefore, in what follows, we focus our quasinormal-mode (QNM) analysis on the \(f(\rho)_{-}\) branch.
	We systematically investigate how angular momentum $l$ and the mass $M$ influence on effective potential, QNMs, and the late-time tail under scalar perturbations. We find that when angular momentum $l$ increases, the peak of the effective potential increases, and the quasinormal frequencies become longer-lived and dominate the time-domain signal. When the mass of black hole $M$ increases, the peak of the effective potential decreases. The late-time tails of black holes with different masses are similar and follow an inverse power-law behavior. To gain a better understanding of the differences in the structure of spacetime for the Schwarzschild black hole, the black hole in the qOS model, and the black hole in the qOS model with Lorentz term, we compared their effective potential and found a small deformation near the even horizon. Furthermore, we compared their time-domain profiles. The profile of the black hole in the qOS model closely matches that of the Schwarzschild black hole, while the black hole in the qOS model with Lorentz term exhibits a slower exponential decay rate. All three configurations display universal inverse power-law tails behavior across all mass configurations.
	
	To better understand the differences in their spacetime structures near the event horizon, we further analyzed their higher overtones and found significant discrepancies in the overtone frequencies, which means the quantum gravity corrections lead to the outburst in the overtones, and this quantum gravity corrections also different from the fundamental mode. We further investigated the quantum-correction parameter $\alpha$, how it affects the higher overtones, we observe that for $l =1$ and $l =2$, $Re(\omega)$ increases monotonically with increasing $\alpha$, except for the case $n=3$, where a non-monotonic behavior is observed. Meanwhile, $Im(\omega)$ decreases monotonically with increasing $\alpha$ for both $l =1$ and $l =2$.
	
	The cosmological constant is generally believed to be the simplest explanation for the accelerated expansion of our Universe, while a negative cosmological constant introduces rich and distinctive physical signatures. 
	Therefore, we consider the vacuum exterior metrics with a negative cosmological constant for both the $f(\rho)_{-}$ and $f(\rho)_{+}$ branches, and investigate their thermodynamic properties.
	
	For the vacuum exterior metric corresponding to the $f(\rho)_{-}$ branch with a negative cosmological constant, we find that for small $r_{h}$, unlike the classical case where the Hawking temperature decreases monotonically from infinity, the temperature in the LQG-inspired black hole initially increases from zero, reaches a maximum, and then decreases. As $r_{h}$ becomes large, the behavior of the LQG-inspired black hole becomes similar to the classical scenario. For the heat capacity, we find that the LQG-inspired black hole exhibits an additional phase transition at a small $r_{h}$ compared to the classical scenario. When $r_{h}$ becomes large, the behavior of the LQG-inspired black hole coincides with that of the classical case. We also investigate the entropy of the LQG-inspired black hole. Compared to the classical case, the entropy-to-area ratio $\frac{S}{A}$ of the LQG-inspired black hole initially increases with $r_{h}$, reaches a maximum, and then decreases, eventually converging to the classical case at large $r_{h}$. This indicates that besides the classical area law \(S = \frac{A}{4}\) term, the LQG-inspired black hole also has an additional logarithmic correction term due to the quantum gravity effect.
	
	For the vacuum exterior metric corresponding to the $f(\rho)_{+}$ branch, we first consider the spacetime without a cosmological constant and compare its thermodynamic behavior with that of the classical Schwarzschild case. Since the quantum geometric effects in this model are equivalent to a positive effective cosmological constant $\Lambda_{eff}$, this comparison allows us to examine how such effects influence the thermodynamic properties.
	We find that for the LQG-inspired black hole, the Hawking temperature first decreases with increasing horizon radius $r_{c}$, and then increases linearly, while the Hawking temperature of de Sitter Schwarzschild black hole monotonically increases with $r_{c}$. However, the rate of increase differs between the two cases. The heat capacity of the LQG-inspired black hole exhibits an additional phase transition at small $r_{h}$ compared to the de Sitter Schwarzschild black hole. And at large horizon radius $r_{c}$, the heat capacity of the LQG-inspired black hole becomes negative, indicating that both black holes are thermodynamically unstable.
	Furthermore, as discussed above, even when a negative cosmological constant is introduced into the metric, provided that its absolute value does not exceed the effective cosmological constant $\Lambda_{eff}$, our main conclusions remain unchanged.
	
	\begin{acknowledgements}
		This work is supported by National Natural Science Foundation of China (NSFC) with Grant No. 12275087. 
	\end{acknowledgements}

	\bibliographystyle{unsrt}
	\bibliography{Ref}

\begin{thebibliography}{10}

\bibitem{thiemann2001introduction}
Thomas Thiemann.
\newblock Introduction to modern canonical quantum general relativity.
\newblock {\em arXiv preprint gr-qc/0110034}, 2001.

\bibitem{ashtekar2004background}
Abhay Ashtekar and Jerzy Lewandowski.
\newblock Background independent quantum gravity: a status report.
\newblock {\em Classical and Quantum Gravity}, 21(15):R53, 2004.

\bibitem{han2007fundamental}
Muxin Han, Yongge Ma, and Weiming Huang.
\newblock Fundamental structure of loop quantum gravity.
\newblock {\em International Journal of Modern Physics D}, 16(09):1397--1474,
  2007.

\bibitem{rovelli2004quantum}
Carlo Rovelli.
\newblock {\em Quantum gravity}.
\newblock Cambridge university press, 2004.

\bibitem{song2022thermodynamics}
Shupeng Song, Gaoping Long, Cong Zhang, and Xiangdong Zhang.
\newblock Thermodynamics of isolated horizons in loop quantum gravity.
\newblock {\em Physical Review D}, 106(12):126007, 2022.

\bibitem{rovelli1996black}
Carlo Rovelli.
\newblock Black hole entropy from loop quantum gravity.
\newblock {\em Physical Review Letters}, 77(16):3288, 1996.

\bibitem{meissner2004black}
Krzysztof~A Meissner.
\newblock Black-hole entropy in loop quantum gravity.
\newblock {\em Classical and Quantum Gravity}, 21(22):5245, 2004.

\bibitem{bojowald2008loop}
Martin Bojowald.
\newblock Loop quantum cosmology.
\newblock {\em Living Reviews in Relativity}, 11:1--131, 2008.

\bibitem{ashtekar2011loop}
Abhay Ashtekar and Parampreet Singh.
\newblock Loop quantum cosmology: a status report.
\newblock {\em Classical and Quantum Gravity}, 28(21):213001, 2011.

\bibitem{agullo2014loop}
Ivan Agullo and Alejandro Corichi.
\newblock Loop quantum cosmology.
\newblock In {\em Springer Handbook of Spacetime}, pages 809--839. Springer,
  2014.

\bibitem{ashtekar2006quantum}
Abhay Ashtekar, Tomasz Pawlowski, and Parampreet Singh.
\newblock Quantum nature of the big bang.
\newblock {\em Physical review letters}, 96(14):141301, 2006.

\bibitem{ashtekar2006quantumanalytical}
Abhay Ashtekar, Tomasz Pawlowski, and Parampreet Singh.
\newblock Quantum nature of the big bang: an analytical and numerical
  investigation.
\newblock {\em Physical Review D—Particles, Fields, Gravitation, and
  Cosmology}, 73(12):124038, 2006.

\bibitem{ashtekar2006quantumdynamics}
Abhay Ashtekar, Tomasz Pawlowski, and Parampreet Singh.
\newblock Quantum nature of the big bang: Improved dynamics.
\newblock {\em Physical Review D—Particles, Fields, Gravitation, and
  Cosmology}, 74(8):084003, 2006.

\bibitem{ashtekar2008robustness}
Abhay Ashtekar, Alejandro Corichi, and Parampreet Singh.
\newblock Robustness of key features of loop quantum cosmology.
\newblock {\em Physical Review D—Particles, Fields, Gravitation, and
  Cosmology}, 77(2):024046, 2008.

\bibitem{yang2009alternative}
Jinsong Yang, You Ding, and Yongge Ma.
\newblock Alternative quantization of the hamiltonian in loop quantum
  cosmology.
\newblock {\em Physics Letters B}, 682(1):1--7, 2009.

\bibitem{zhang2016higher}
Xiangdong Zhang.
\newblock Higher dimensional loop quantum cosmology.
\newblock {\em The European Physical Journal C}, 76(7):1--11, 2016.

\bibitem{oppenheimer1939continued}
J~Robert Oppenheimer and Hartland Snyder.
\newblock On continued gravitational contraction.
\newblock {\em Physical Review}, 56(5):455, 1939.

\bibitem{lewandowski2023quantum}
Jerzy Lewandowski, Yongge Ma, Jinsong Yang, and Cong Zhang.
\newblock Quantum oppenheimer-snyder and swiss cheese models.
\newblock {\em Physical Review Letters}, 130(10):101501, 2023.

\bibitem{shi2024higher}
Zijian Shi, Xiangdong Zhang, and Yongge Ma.
\newblock Higher-dimensional quantum oppenheimer-snyder model.
\newblock {\em Physical Review D}, 110(10):104074, 2024.

\bibitem{gong2024quasinormal}
Huajie Gong, Shulan Li, Dan Zhang, Guoyang Fu, and Jian-Pin Wu.
\newblock Quasinormal modes of quantum-corrected black holes.
\newblock {\em Physical Review D}, 110(4):044040, 2024.

\bibitem{yang2023shadow}
Jinsong Yang, Cong Zhang, and Yongge Ma.
\newblock Shadow and stability of quantum-corrected black holes.
\newblock {\em The European Physical Journal C}, 83(7):619, 2023.

\bibitem{zhang2023loop}
Xiangdong Zhang.
\newblock Loop quantum black hole.
\newblock {\em Universe}, 9(7):313, 2023.

\bibitem{assanioussi2018emergent}
Mehdi Assanioussi, Andrea Dapor, Klaus Liegener, and Tomasz Paw{\l}owski.
\newblock Emergent de sitter epoch of the quantum cosmos from loop quantum
  cosmology.
\newblock {\em Physical review letters}, 121(8):081303, 2018.

\bibitem{assanioussi2019emergent}
Mehdi Assanioussi, Andrea Dapor, Klaus Liegener, and Tomasz Paw{\l}owski.
\newblock Emergent de sitter epoch of the loop quantum cosmos: A detailed
  analysis.
\newblock {\em Physical Review D}, 100(8):084003, 2019.

\bibitem{han2020effective}
Muxin Han and Hongguang Liu.
\newblock Effective dynamics from coherent state path integral of full loop
  quantum gravity.
\newblock {\em Physical Review D}, 101(4):046003, 2020.

\bibitem{yang2023alternative}
Jinsong Yang, Cong Zhang, and Xiangdong Zhang.
\newblock Alternative k=-1 loop quantum cosmology.
\newblock {\em Physical Review D}, 107(4):046012, 2023.

\bibitem{long2021energy}
Gaoping Long, Yunlong Liu, and Xiangdong Zhang.
\newblock Energy conditions in the new model of loop quantum cosmology.
\newblock {\em Chinese Physics C}, 45(11):115102, 2021.

\bibitem{li2018towards}
Bao-Fei Li, Parampreet Singh, and Anzhong Wang.
\newblock Towards cosmological dynamics from loop quantum gravity.
\newblock {\em Physical Review D}, 97(8):084029, 2018.

\bibitem{zhang2021loop}
Xiangdong Zhang, Gaoping Long, and Yongge Ma.
\newblock Loop quantum gravity and cosmological constant.
\newblock {\em Physics Letters B}, 823:136770, 2021.

\bibitem{lin2024quasinormal}
Jianhui Lin, Mois{\'e}s Bravo-Gaete, and Xiangdong Zhang.
\newblock Quasinormal modes, greybody factors, and thermodynamics of four
  dimensional ads black holes in critical gravity.
\newblock {\em Physical Review D}, 109(10):104039, 2024.

\bibitem{abbott2017gw170817}
Benjamin~P Abbott, Rich Abbott, TDea Abbott, Fausto Acernese, Kendall Ackley,
  Carl Adams, Thomas Adams, Paolo Addesso, Rana~X Adhikari, Vaishali~B Adya,
  et~al.
\newblock Gw170817: observation of gravitational waves from a binary neutron
  star inspiral.
\newblock {\em Physical review letters}, 119(16):161101, 2017.

\bibitem{abbott2017gravitational}
Benjamin~P Abbott, Robert Abbott, TD~Abbott, F~Acernese, K~Ackley, C~Adams,
  T~Adams, P~Addesso, RX~Adhikari, VB~Adya, et~al.
\newblock Gravitational waves and gamma-rays from a binary neutron star merger:
  Gw170817 and grb 170817a.
\newblock {\em The Astrophysical Journal Letters}, 848(2):L13, 2017.

\bibitem{giesler2019black}
Matthew Giesler, Maximiliano Isi, Mark~A Scheel, and Saul~A Teukolsky.
\newblock Black hole ringdown: the importance of overtones.
\newblock {\em Physical Review X}, 9(4):041060, 2019.

\bibitem{oshita2021ease}
Naritaka Oshita.
\newblock Ease of excitation of black hole ringing: Quantifying the importance
  of overtones by the excitation factors.
\newblock {\em Physical Review D}, 104(12):124032, 2021.

\bibitem{forteza2021high}
Xisco~Jim{\'e}nez Forteza and Pierre Mourier.
\newblock High-overtone fits to numerical relativity ringdowns: Beyond the
  dismissed n= 8 special tone.
\newblock {\em Physical Review D}, 104(12):124072, 2021.

\bibitem{shao2024scalar}
Cai-Ying Shao, Cong Zhang, Wei Zhang, and Cheng-Gang Shao.
\newblock Scalar fields around a loop quantum gravity black hole in de sitter
  spacetime: Quasinormal modes, late-time tails and strong cosmic censorship.
\newblock {\em Physical Review D}, 109(6):064012, 2024.

\bibitem{cardoso2018quasinormal}
Vitor Cardoso, Joao~L Costa, Kyriakos Destounis, Peter Hintz, and Aron Jansen.
\newblock Quasinormal modes and strong cosmic censorship.
\newblock {\em Physical review letters}, 120(3):031103, 2018.

\bibitem{poisson1990internal}
Eric Poisson and Werner Israel.
\newblock Internal structure of black holes.
\newblock {\em Physical Review D}, 41(6):1796, 1990.

\bibitem{hawking1974black}
Stephen~W Hawking.
\newblock Black hole explosions?
\newblock {\em Nature}, 248(5443):30--31, 1974.

\bibitem{hawking1975particle}
Stephen~W Hawking.
\newblock Particle creation by black holes.
\newblock {\em Communications in mathematical physics}, 43(3):199--220, 1975.

\bibitem{bardeen1973four}
James~M Bardeen, Brandon Carter, and Stephen~W Hawking.
\newblock The four laws of black hole mechanics.
\newblock {\em Communications in mathematical physics}, 31:161--170, 1973.

\bibitem{bekenstein1973black}
Jacob~D Bekenstein.
\newblock Black holes and entropy.
\newblock {\em Physical Review D}, 7(8):2333, 1973.

\bibitem{weinberg1989cosmological}
Steven Weinberg.
\newblock The cosmological constant problem.
\newblock {\em Reviews of modern physics}, 61(1):1, 1989.

\bibitem{peebles2003cosmological}
P~James~E Peebles and Bharat Ratra.
\newblock The cosmological constant and dark energy.
\newblock {\em Reviews of modern physics}, 75(2):559, 2003.

\bibitem{du2023topological}
Yongbin Du and Xiangdong Zhang.
\newblock Topological classes of black holes in de-sitter spacetime.
\newblock {\em The European Physical Journal C}, 83(10):927, 2023.

\bibitem{sekiwa2006thermodynamics}
Yuichi Sekiwa.
\newblock Thermodynamics of de sitter black holes: Thermal cosmological
  constant.
\newblock {\em Physical Review D—Particles, Fields, Gravitation, and
  Cosmology}, 73(8):084009, 2006.

\bibitem{kubizvnak2012p}
David Kubiz{\v{n}}{\'a}k and Robert~B Mann.
\newblock P- v criticality of charged ads black holes.
\newblock {\em Journal of High Energy Physics}, 2012(7):1--25, 2012.

\bibitem{dayyani2018critical}
Z~Dayyani, A~Sheykhi, MH~Dehghani, and S~Hajkhalili.
\newblock Critical behavior and phase transition of dilaton black holes with
  nonlinear electrodynamics.
\newblock {\em The European Physical Journal C}, 78:1--19, 2018.

\bibitem{wei2020extended}
Shao-Wen Wei and Yu-Xiao Liu.
\newblock Extended thermodynamics and microstructures of four-dimensional
  charged gauss-bonnet black hole in ads space.
\newblock {\em Physical Review D}, 101(10):104018, 2020.

\bibitem{hawking1983thermodynamics}
Stephen~W Hawking and Don~N Page.
\newblock Thermodynamics of black holes in anti-de sitter space.
\newblock {\em Communications in Mathematical Physics}, 87:577--588, 1983.

\bibitem{darmois1927equations}
Georges Darmois.
\newblock {\em Les {\'e}quations de la gravitation einsteinienne}.
\newblock Number~25. Gauthier-Villars et. cie., 1927.

\bibitem{israel1966singular}
Werner Israel.
\newblock Singular hypersurfaces and thin shells in general relativity.
\newblock {\em Il Nuovo Cimento B (1965-1970)}, 44(1):1--14, 1966.

\bibitem{markovic2000gravitational}
Dragoljub Markovic and Stuart~L Shapiro.
\newblock Gravitational collapse with a cosmological constant.
\newblock {\em Physical Review D}, 61(8):084029, 2000.

\bibitem{papanikolaou2023primordial}
Theodoros Papanikolaou.
\newblock Primordial black holes in loop quantum cosmology: the effect on the
  threshold.
\newblock {\em Classical and Quantum Gravity}, 40(13):134001, 2023.

\bibitem{schutz1985black}
Bernard~F Schutz and Clifford~M Will.
\newblock Black hole normal modes: a semianalytic approach.
\newblock {\em The Astrophysical Journal}, 291:L33--L36, 1985.

\bibitem{iyer1987black}
Sai Iyer and Clifford~M Will.
\newblock Black-hole normal modes: A wkb approach. i. foundations and
  application of a higher-order wkb analysis of potential-barrier scattering.
\newblock {\em Physical Review D}, 35(12):3621, 1987.

\bibitem{konoplya2003quasinormal}
RA~Konoplya.
\newblock Quasinormal behavior of the d-dimensional schwarzschild black hole
  and the higher order wkb approach.
\newblock {\em Physical Review D}, 68(2):024018, 2003.

\bibitem{matyjasek2017quasinormal}
Jerzy Matyjasek and Micha{\l} Opala.
\newblock Quasinormal modes of black holes: The improved semianalytic approach.
\newblock {\em Physical Review D}, 96(2):024011, 2017.

\bibitem{konoplya2019higher}
RA~Konoplya, A~Zhidenko, and AF~Zinhailo.
\newblock Higher order wkb formula for quasinormal modes and grey-body factors:
  recipes for quick and accurate calculations.
\newblock {\em Classical and Quantum Gravity}, 36(15):155002, 2019.

\bibitem{abbott2016binary}
Benjamin~P Abbott, R~Abbott, TD~Abbott, MR~Abernathy, Fausto Acernese,
  K~Ackley, C~Adams, T~Adams, Paolo Addesso, RX~Adhikari, et~al.
\newblock Binary black hole mergers in the first advanced ligo observing run.
\newblock {\em Physical Review X}, 6(4):041015, 2016.

\bibitem{konoplya2024first}
RA~Konoplya and A~Zhidenko.
\newblock First few overtones probe the event horizon geometry.
\newblock {\em Journal of High Energy Astrophysics}, 44:419--426, 2024.

\bibitem{oshita2023slowly}
Naritaka Oshita and Daichi Tsuna.
\newblock Slowly decaying ringdown of a rapidly spinning black hole: Probing
  the no-hair theorem by small mass-ratio mergers with lisa.
\newblock {\em Physical Review D}, 108(10):104031, 2023.

\bibitem{konoplya2023quasinormal}
RA~Konoplya.
\newblock Quasinormal modes in higher-derivative gravity: Testing the black
  hole parametrization and sensitivity of overtones.
\newblock {\em Physical Review D}, 107(6):064039, 2023.

\bibitem{berti2003asymptotic}
Emanuele Berti and Kostas~D Kokkotas.
\newblock Asymptotic quasinormal modes of reissner-nordstr{\"o}m and kerr black
  holes.
\newblock {\em Physical Review D}, 68(4):044027, 2003.

\bibitem{konoplya2023bardeen}
RA~Konoplya, D~Ovchinnikov, and B~Ahmedov.
\newblock Bardeen spacetime as a quantum corrected schwarzschild black hole:
  Quasinormal modes and hawking radiation.
\newblock {\em Physical Review D}, 108(10):104054, 2023.

\bibitem{lin2024effective}
Jianhui Lin and Xiangdong Zhang.
\newblock Effective four-dimensional loop quantum black hole with a
  cosmological constant.
\newblock {\em Physical Review D}, 110(2):026002, 2024.

\bibitem{wang2024thermodynamics}
Rui-Bo Wang, Shi-Jie Ma, Lei You, Yu-Cheng Tang, Yu-Hang Feng, Xian-Ru Hu, and
  Jian-Bo Deng.
\newblock Thermodynamics of ads-schwarzschild-like black hole in loop quantum
  gravity.
\newblock {\em The European Physical Journal C}, 84(11):1161, 2024.

\bibitem{bentivegna2008anti}
Eloisa Bentivegna and Tomasz Pawlowski.
\newblock Anti-de sitter universe dynamics in loop quantum cosmology.
\newblock {\em Physical Review D—Particles, Fields, Gravitation, and
  Cosmology}, 77(12):124025, 2008.

\bibitem{huang2023novel}
Yang Huang, Dao-Jun Liu, and Hongsheng Zhang.
\newblock Novel black holes in higher derivative gravity.
\newblock {\em Journal of High Energy Physics}, 2023(2):1--11, 2023.

\bibitem{cai2002gauss}
Rong-Gen Cai.
\newblock Gauss-bonnet black holes in ads spaces.
\newblock {\em Physical Review D}, 65(8):084014, 2002.

\bibitem{kubizvnak2015black}
David Kubiz{\v{n}}{\'a}k and Robert~B Mann.
\newblock Black hole chemistry.
\newblock {\em Canadian Journal of Physics}, 93(9):999--1002, 2015.

\bibitem{kubizvnak2017blackLambda}
David Kubiz{\v{n}}{\'a}k, Robert~B Mann, and Mae Teo.
\newblock Black hole chemistry: thermodynamics with lambda.
\newblock {\em Classical and Quantum Gravity}, 34(6):063001, 2017.

\bibitem{dolan2011pressure}
Brian~P Dolan.
\newblock Pressure and volume in the first law of black hole thermodynamics.
\newblock {\em Classical and Quantum Gravity}, 28(23):235017, 2011.

\bibitem{diaz2012isolated}
Jacobo Diaz-Polo, Daniele Pranzetti, et~al.
\newblock Isolated horizons and black hole entropy in loop quantum gravity.
\newblock {\em SIGMA. Symmetry, Integrability and Geometry: Methods and
  Applications}, 8:048, 2012.

\bibitem{frodden2012black}
Ernesto Frodden, Amit Ghosh, and Alejandro Perez.
\newblock Black hole entropy in lqg: Recent developments.
\newblock In {\em AIP Conference Proceedings}, volume 1458, pages 100--115.
  American Institute of Physics, 2012.

\end{thebibliography}

\end{document}